\documentstyle[12pt]{article}

\begin{document}
\input epsf.tex
\pagestyle{empty}
\begin{flushright}
BGU PH-95/05\\CERN-TH/95-147
\end{flushright} 
\vspace{.5in}
\begin{large}
\begin{center} {\bf DILATON-DRIVEN INFLATION \\
 IN STRING COSMOLOGY} \end{center}
\end{large}

\noindent
\begin{center} Ram Brustein $^{*,**)}$
\\ {\ }\\ Theory Division, CERN 
CH-1211 Geneva 23, Switzerland
\end{center}
\vspace{1in}
\begin{center} ABSTRACT \end{center}
I present an outline for cosmological evolution in
the framework of string theory with emphasis  on a phase of
dilaton-driven kinetic inflation. It is shown that a typical background of  stochastic
 gravitational radiation is generated, with strength that may allow its detection in  
future gravity wave experiments.
\vspace{.5in}
\begin{center} 
$^{*)}$Present address:
Department of Physics, Ben-Gurion
University, \\ Beer-Sheva 84105, Israel.\\
$^{**)}$Contribution to the proceedings of  the International conference\\  
Unified symmetry in the small and in the large,\\
 February 2-5, 1995, Coral Gables , Florida. \end{center}  
\noindent \rule[.1in]{16.5cm}{.002in}\\
\begin{flushleft} CERN-TH/95-147 \\ June 1994
\end{flushleft} \vfill\eject 

\setcounter{page}{1}
\pagestyle{plain}

{\bf INTRODUCTION}\\

I present an outline of cosmological evolution in
the framework of string theory. The main emphasis is on a phase of
dilaton-driven kinetic inflation and its possible observable
consequences, in particular, a background of  stochastic
 gravitational radiation.  The results concerning the produced
spectrum of gravitational radiation were obtained in
\cite{bggmv,bggv}. More details on various aspects of the suggested
outline and  additional  references may be found in [3-11].
\centerline{ \ } \\ \centerline{ \  } \\
{\bf POTENTIAL-DRIVEN INFLATION}\\

Inflationary evolution of the universe requires a source of
energy to drive the expansion.  The conventional
expectation is that the energy source is dominated by potential
energy of scalar fields, called  inflatons \cite{inflation}.  The
inflatons  are expected to posses non-vanishing potential energy
during some  phase in  their evolution in which inflationary
expansion  takes place. Eventually, the inflatons settle down to the
true minimum  of their potential where the potential energy vanishes,
thus depriving the universe of the necessary source to drive its
accelerated  expansion. The inflationary phase ends and the universe
continues to  expand sub-luminally until today. If one tries to
implement  similar ideas in the framework of string theory, an
apparent problem is immediately encountered \cite{bs,clo}. String
theory does indeed  contain many
scalar fields, called moduli, which seem particularly suitable for
the job of inflatons \cite{moduli,bbmss}.
 Among the moduli the dilaton $\phi$ is an
important and universal field whose expectation value determines
the string coupling parameter $g_s^2\sim\langle exp(\phi)\rangle$.
It couples to all other fields with
gravitational strength. If some scalar field, for example,  one of
the moduli fields acquires a non-vanishing potential  so does the
dilaton. The type of generated dilaton potential depends on the
details of the model. Two types are distinguished, perturbative
$V(\phi)\sim exp({-\alpha\phi/M_{Pl}})$, and  non perturbative
$V(\phi)\sim exp (-exp({-\beta\phi/M_{Pl}}))$, with particular
numerical parameters $\alpha$, $\beta$. The equations of motion for
the resulting string dilaton-gravity,  assuming  isotropic and
homogeneous universe \begin{eqnarray}  ds^2 &=& -dt^2+a^2(t) dx_i
dx^i\nonumber \\ \phi&=&\phi(t),  \end{eqnarray}  are the following
\begin{eqnarray}
H^2 &=& \frac{8\pi}{3 M_{Pl}^2}\left(\frac{1}{2} \dot\phi^2+
V(\phi)\right) \nonumber \\
\ddot\phi +3 H \dot\phi &=& -\frac{ d V}{d\phi}
\label{dilgravpot}
\end{eqnarray}
The Hubble parameter, $H$, is related to  the scale factor,
$a$ in the usual way,
$H\equiv\frac{\hbox{\large $\dot a$}}{\hbox{\large $a$}}$ and $V$
is the potential.  Consider, for example, the (unrealistic) case of
exponential  potential
$V=V_0\ exp ({-\alpha\phi/M_{Pl}})$ for which can solve eqs.
(\ref{dilgravpot}) explicitly  \begin{equation}
a(t)=a_0\ t^{\hbox{${16\pi}/{\alpha^2}$}}.
\label{dilsol}
\end{equation}
If the potential is steeper than the critical steepness $\alpha=4
\sqrt{\pi}$,  the dilaton kinetic energy becomes dominant over
potential energy and the expansion is subluminal. The generic
situation in  string theory is that the potentials in several models
are steeper than critical and therefore  potential-driven  inflation
requires  special situations and is generally speaking hard to
obtain. Recently, some progress has been made towards characterizing
requirements from models in which potential-driven inflation could
be  supported \cite{bbmss,dv}.  \ \centerline{ \ } 
\centerline{ \
} \\ {\bf DILATON-DRIVEN KINETIC INFLATION}\\

The outline for cosmological evolution   that I present here relies
heavily on the fact that the kinetic energy of the dilaton tends to
dominate the energy density. Instead of trying to fight  this
tendency,  one accepts it and
 turns this feature into a virtue, using it to drive kinetic energy
dominated inflationary evolution. Kinetic inflation was also
discussed in \cite{LKI}. The evolution starts when the  dilaton is
deep in the weak-coupling region ($\phi\ll-1$) and Hubble parameter,
$H$, is small. The evolution in this epoch is shown below to be
accelerated expansion dominated
by the dilaton kinetic energy and  determined by the vacuum solution
of the string dilaton-gravity equations of motion \cite{gv1}.
To describe the first phase in more detail, look for solutions of
the effective
string  equations of motion in which the metric  is  of the
isotropic, FRW type with  vanishing spatial curvature  and the
dilaton depends only on time.  One finds three independent first
order equations for the   dilaton and  $H$
\renewcommand{\theequation}{\arabic{equation}{a}} \begin{equation}
\dot H=\pm H\sqrt{3H^2+U + \hbox{\large $e^\phi$} \rho}
-\frac{1}{2} U' +
\frac{1}{2} \hbox{\large $e^\phi$} p \setcounter{equation}{4}
\end{equation}
\renewcommand{\theequation}{\arabic{equation}{b}}\begin{equation}
\setcounter{equation}{4} \dot \phi= 3H\pm\sqrt{3H^2+U+\hbox{\large
$e^\phi$} \rho}  \end{equation}
\renewcommand{\theequation}{\arabic{equation}{c}} \begin{equation}
\dot \rho + 3 H (\rho + p) = 0
\setcounter{equation}{4}\label{fstordm}   \end{equation}
\renewcommand{\theequation}{\arabic{equation}}  where $U= e^{\phi}
V$. Some sources in the form of an ideal fluid were included
\cite{gv1} as well. The ($\pm$) signifies that either  $(+)$  or
$(-)$ is chosen for both  equations simultaneously.  The solutions
of equations (4a-4c) belong to two branches, according  to which sign
is chosen.  In the absence  of any potential or sources the $(+)$
branch solution for $\{H,\phi\}$    is  given by
 \begin{eqnarray}
 H^{(+)}&=&\pm \frac{1}{\sqrt{3}}\frac{1}{t-t_0}   \nonumber \\
\phi^{(+)}&=&\phi_0 + (\pm\sqrt{3}-1)\ln({t_0 -t}) ~~, ~~ t<t_0
\label{plusol}
\end{eqnarray}
This solution describes either  accelerated contraction  and
evolution towards weak coupling or accelerated inflationary
expansion  and  evolution from  a cold,  flat and weakly coupled
universe towards a  hot, curved and strongly coupled one.  I assume
that the initial conditions are such that the latter is  chosen. In
general, the effects of a potential and sources on this branch are
quite mild. After a period of time, of length  determined by the
initial conditions,  a  ``Branch Change" event from the
dilaton-driven accelerated expansion era into what  will  eventually
become a phase of decelerated expansion  has to occur. It occurs
either when curvatures and kinetic  energies reach the string
curvature or when quantum effects become strong enough.   The correct
dynamical description of this phase  should, therefore, be stringy in
nature. If the value of the dilaton is small  throughout this stage
of evolution, dynamics can be described by classical string theory
in terms of a two-dimensional conformal field theory. This stage is
not yet  well
understood. At the moment, the only  existing examples are not
quite realistic
 \cite{kk,ts}. More ideas about this stage may be found in
\cite{martinec,DM}.  The value which the dilaton takes at the end of
this epoch $\phi_{end}$
 is  an important
parameter.  After the ``Branch Change" event, the universe cools
down and may  be described accurately, again, by means of string
dilaton-gravity  effective theory. Now, however, radiation and matter
are important factors. The dilaton remains approximately  at the
value $\phi_{end}$.  The universe evolves as a regular
Friedman-Robertson-Walker  (FRW) radiation-dominated universe.
\centerline{ \ } \\ \centerline{ \  } \\ {\bf TENSOR PERTURBATIONS
AND RELIC GRAVITATIONAL WAVES}\\

The phase of accelerated evolution, described in the previous section, produces
a typical and unique spectrum of gravitationl radiation.  The basic
mechanism is by now well known \cite{first} (see \cite{MFB, LL} for
recent reviews).  Quantum mechanical perturbations exist as tiny
wrinkles on top of the classically homogeneous  and isotropic
background. These wrinkles are then magnified by the accelerated
evolution and become classical stochastic inhomogeneities. Below I
sketch the derivation of the spectrum of tensor perturbations. Many
technical elements are omitted here and can be found in gory details
in \cite{bggmv}. The classical solution  (\ref{plusol}) in conformal
time $\eta$, where $dt\equiv a d\eta$, is given by
\begin{equation}
g_{\mu\nu}= diag(a^2(\eta),-a^2(\eta)\delta_{ij})\ \ \ i,j=1,2,3
\end{equation}
where
\begin{equation}
a^2(\eta)\sim |\eta|^{1/2}, \ \ \
\phi(\eta)\sim-\sqrt{3}\ln|\eta|+\phi_0
\end{equation}
for $\eta\rightarrow 0_-$.
One expands the metric around the
classical solution $g=g_{cs}+\delta g$ where $g_{cs}$ is given in
the previous equation and
$\delta g_{ij}=-a^2(\eta) h_{ij}(\eta,\vec x)$. The resulting
equation of motion  for
each of the two independent tensor perturbation components is given in
Fourier space by
\begin{equation}
h''_k+2 {{a'}\over{a}} h'_k+k^2 h_k=0
\end{equation}
and has the general solution
$h_k=A_k+B_k\ln|k\eta|$. Initial conditions corresponding to
quantum fluctuations at short scales
$h_k \sim {1}/({a \sqrt{k}})\ exp[{i( \vec k\cdot \vec x- k\eta)}]$,
determine $h_k$
\begin{equation}
|h_k|\simeq {\ln|k\eta|\over \sqrt{k} a_{HC}}\simeq \ln|k\eta|.
\label{hk}
\end{equation}
The amplitude of stochastic tensor perturbations in $x$
space is characterized  by $|\delta h_k|\sim k^{3/2} |h_k|$.
From eq.(\ref{hk}) we obtain
\begin{equation}
|\delta h_k|^2(\eta)\sim \left({H_{max}\over M_{Pl}}\right)^2
|k \eta_{max}|^3 \left(\ln|k\eta|\right)^2.
\end{equation}
The end of the dilaton-driven epoch is assumed \cite {BV} to
take place when the curvature scale $H$ reaches the string scale
$M_s$. In the Einstein frame, in which $M_{Pl}$ is constant, the string
scale depends upon the dilaton as $M_s=\exp(\phi/2) M_{Pl}$. Thus we
assume 
 the dilaton-driven era to end at conformal time $|\eta| = \eta_1$ where
$ H_1 \simeq (\eta_1 a(\eta_1))^{-1} = M_s(\eta_1) =
 \exp(\phi(\eta_1)/2) M_{Pl}$.
At the end of the dilaton-driven era we thus have
\begin{equation}
\left|\delta {h_k}(\eta_{1})\right| \sim
 {H_1\over M_{Pl}} (k\eta_1)^{3/2} \ln(k\eta_1)
\label{hend}
\end{equation}
This is the final result for the  primordial
spectrum of tensor perturbations. From the primordial spectrum  one
wishes to compute the observable spectrum today.  A nice feature of
gravitational waves is that gravitons are affected  practically  only
by the evolution of the  background curvature since  right after the
``Branch Change" era. Thus the spectrum that should be seen today
should mainly reflect what happened in the very early universe
processed through presumed  known background evolution. While
frequencies shift according to the evolution of the background scale
factor throughout the evolution, amplitudes of tensor perturbations
freeze while outside the horizon and evolve only when inside the
horizon. If the dilaton-driven era is followed by a stringy phase
characterized by an almost constant value of $H$, we expect  scales
which went out of the horizon during the dilaton-driven era to keep
moving further
 outside and to reenter only much later,
 during the radiation, or possibly even  matter dominated era.
If we assume this to be the case for all (comoving) scales
larger than $\eta_1^{-1}$, we must also assume that
$h_k$ remains frozen, for all these scales,
 at the value given in eq.(\ref{hend}) until reentry.

The result is \cite{bggv} that the part of the processed  spectrum
which lies below a certain maximal  frequency $\omega_{max}$,  the
highest frequency  amplified during the dilaton-driven
era,
  is presently given, in the string frame,  by
\begin{equation}
|\delta {h_{\omega}}| = \sqrt{H_0 / M_s} z_{eq}^{-1/4} z_{out}^{1/2}
\exp{(\frac{1}{2} \phi_{end})}\;
({\omega\over \omega_{max}})^{\frac{1}{2}} \ln ({\omega \over
\omega_{max}}) \label{deltah}
\end{equation}
where $z_{out}(k) = a_{re}(k)/a_{ex}(k)$ is the
red-shift
while the scale $k^{-1}$ was outside the horizon, $z_{eq}$ is the
red-shift from the matter-radiation equality epoch until today,
$M_s$ is the present
value of string scale (usually estimated to be about
$2-5 \cdot 10^{17} GeV$), $H_0\sim 10^{-18}Hz$ is the present value  of the
Hubble parameter, $\phi_{end}=\phi(\eta_1)$, and 
\begin{equation}
\omega_{max} =\sqrt{H_0 M_s} z_{eq}^{-1/4} z_{out}^{-1/2}.
\label{omega}
\end{equation}
The fraction of energy in gravitational waves in units of the
critical density is given by
\begin{equation}
{d\Omega \over d \ln \omega} =  z_{eq}^{-1} \exp (\phi_{end})\;
({\omega\over \omega_{max}})^3 \ln^2 ({\omega \over \omega_{max}}).
\label{smega}
\end{equation}
Equations (\ref{deltah}-\ref{smega}) were derived assuming reentry during 
the radiation dominated era and 
should be taken as good estimates and not as numerically accurate
expressions. The processed spectrum of gravitational radiation is
presented graphically in Figure 1,\\

\centerline{\epsfysize=3.5in\epsfbox{cgfig.epsf}}

\noindent
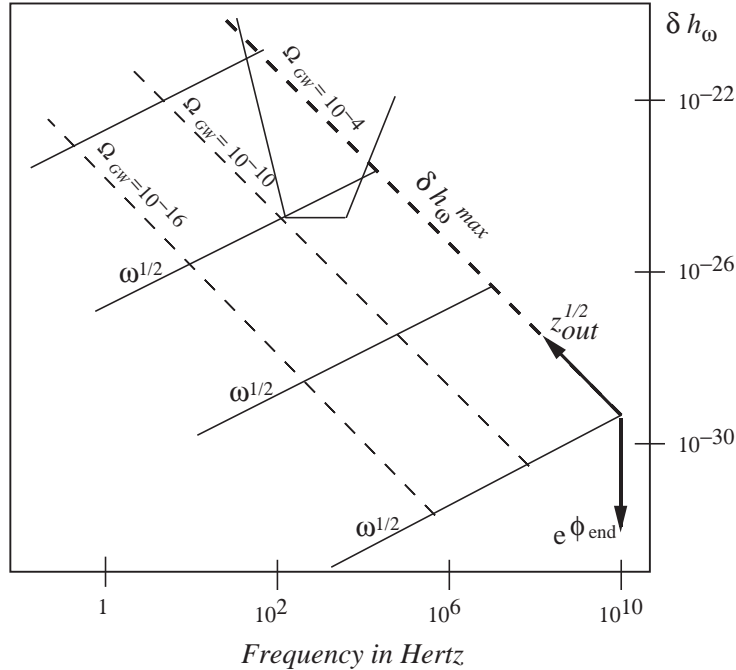
{\small {\bf Figure 1.} The characteristic spectral amplitude
 of gravitational waves $|\delta {h_{\omega}}|$. The solid lines
 show several individual spectra for different values of $z_{out}$
 and $\phi_{end}=0$. The thick dashed line shows the maximum
amplitude $|\delta h_\omega^{max}|$ as a function of $z_{out}$
for $\phi_{end}=0$. The dashed lines are lines of fixed $\phi_{end}$
and therefore lines of constant energy density. $\Omega_{GW}$ is
the maximal amount of gravitational energy density at a given
$\phi_{end}$. Also shown in the figure is a triangular shape marking
 the sensitivity goals for detection of stochastic background
$h_{3/yr}$, of the  ``Advanced LIGO".}

Two possible devices may be able to detect the predicted
stochastic gravitational wave background, in the lower frequency 
region $1-10^4$ Hz, large interferometers, such as
the planned LIGO\cite{LIGO} and VIRGO\cite{VIRGO} and in the higher 
range of frequencies $10^6-10^9$ Hz, room-size microwave cavities.
For a given set of parameters the amplitude grows
as $|\delta {h_{\omega}}|\sim \omega^{1/2}$
and therefore it  may seem that the  best sensitivity
for detection is at the high end of the spectrum
$\omega=\omega_{max}$. However, the noise in a given interferometer
grows as  $h_n\sim \omega^{5/4}$ \cite{thorne}.  Therefore for a
given interferometer the best sensitivity actually is in the lowest
frequency range available.  Microwave cavities may be operated as
gravity wave detectors \cite{grishchuk} for the high frequency range
$10^6-10^9$ Hz. For the MHz range specific suggestions
\cite{picasso,caves} have been implemented \cite{reece}, but not
operated as gravitational radiation detector.
As can be seen from Figure 1, the required
sensitivity for detection at the MHz region is $h_c\sim 10^{-26}$
corresponding to $h_{3/yr}$ of the same order  and therefore to a
noise level of  $h_n\sim 10^{-23}$ \cite{thorne}, assuming a
bandwidth of MHz. With attainable
 $Q$ factors of the order of $10^{11}$, this sensitivity goal does
not seem out  of reach. For the GHz region the  required sensitivity
is $h_c\sim 10^{-28}$   corresponding to $h_n\sim 10^{-24}$.
\noindent

{\ }\centerline{ \ } \\ \centerline{ \  } \\
{\bf ACKNOWLEDGMENT}\\

Research supported in part by an Alon Grant. I would like to thank
M. Gasperini, M. Giovannini, V. Mukhanov and G. Veneziano
for enjoyable and fruitful collaboration and S. Finn, P. Michelson
P. Saulson and  N. Robertson for discussions about gravity wave
detectors. 

\begin{thebibliography}{99}
\bibitem{bggmv} R. Brustein, M. Gasperini, M. Giovannini,
V. Mukhanov and G. Veneziano, ``Metric perturbations in
dilaton-driven inflation", CERN-TH. 7544/94,  Phys. Rev. D (1995), in
Press. \bibitem{bggv} R. Brustein, M. Gasperini, M. Giovannini and
G. Veneziano, CERN preprint.
\bibitem{gv1} G. Veneziano, Phys. Lett. B265 (1991) 287.
\bibitem{dildriv} 
M. Gasperini and G. Veneziano, Astropart. Phys. 1 (1993) 317;
M. Gasperini and G. Veneziano, Mod. Phys. Lett. A8 (1993) 3701;\\
M. Gasperini and G. Veneziano, Phys. Rev. D50 (1994) 2519.
\bibitem{BV} R. Brustein and G. Veneziano, Phys. Lett. B329 (1994) 429.
\bibitem{GAS} M. Gasperini, ``Phenomenological aspects of the
pre-big-bang
scenario in string cosmology", in Proc. of the 2nd Journee
Cosmologie, Paris,
June 1994 (World Scientific P.C., Singapore), Torino University
Preprint DFTT-24/94; \\
M. Gasperini, ``The inflationary role of the dilaton in string
cosmology", in
Proc. of the First Int. Workshop on Birth of the Universe and
Fundamental
Physics, Rome, May 1994 (Springer-Verlag, Berlin), Torino University
Preprint DFTT-29/94.
\bibitem{Ven2} G. Veneziano, ``Strings, Cosmology, ...and a Particle"
talk given at PASCOS '94, preprint CERN-TH.7502/94 (November 1994).
\bibitem{RAMY} R. Brustein,
``The role of the superstring dilaton in cosmology and particle
physics", in Proc. of the XXIX Rencontres de Moriond, Meribel, March
1994;\\
R. Brustein, ``Cosmology and models of supersymmetry breaking in string
theory", Proc. of the SUSY94 Workshop, Ann Arbor, May 1994.
\bibitem{GG1} M. Gasperini and M. Giovannini, Phys. Rev. D47 (1993) 1529.
\bibitem{ggv}
M. Gasperini and M. Giovannini and G. Veneziano,
``Primordial magnetic fileds from string cosmology", preprint  hep-th/9504083.
\bibitem {behrndt } K. Behrndt and S. Forste, Nucl. Phys. B430 (1994) 441; \\
E. J. Copeland, A. Lahiri and  D. Wands, Phys. Rev. D50 (1994) 4880;\\
C. Angelantonj, L. Amendola, M. Litterio and F. Occhionero,
 Phys. Rev. D51 (1995) 1607.
\bibitem{inflation} A. Guth, Phys. Rev. D23 (1981) 347;\\
A. Linde, Phys. Lett 108B (1982) 389;\\
A. Albrecht and P.J. Steinhardt, Phys. Rev. Lett. 48 (1982) 122;\\
A. Linde, Phys. Lett 129B (1983) 177.
\bibitem{bs} R. Brustein and P. J. Steinhardt, Phys. Lett. B302 (1993) 196.
\bibitem{clo} B. Campbell, A. Linde and K. Olive, Nucl. Phys. B335 (1991) 146.
\bibitem{moduli}  P. Binetruy and M.K. Gaillard,
   Phys. Rev. D34 ( 1986) 3069;\\
M. Cvetic and  R. L. Davis, Phys. Lett. B296 (1992) 316;\\
\bibitem{bbmss}
 T. Banks, M. Berkooz, S.H. Shenker, G. Moore and P.J. Steinhardt,
``Modular cosmology",   preprint hep-th/9503114.
\bibitem{dv} T. Damour and A. Vilenkin,
``String theory and inflation", preprint  hep-th/9503149.
\bibitem{LKI} J. Levin,  Phys. Rev. D51 (1995) 1536.
\bibitem{kk}
E. Kiritsis and K. Kounnas, Phys. Lett. B331 (1994) 51; \\
E. Kiritsis and K. Kounnas, in Proc. of the 2nd Journee
Cosmologie, Paris, June 1994 (World Scientific P.C., Singapore).
\bibitem{ts}
A. A. Tseytlin,  Phys. Lett. B334 (1994) 315.
\bibitem{martinec} E. Martinec, Class.Quant.Grav.12 (1995) 941.
\bibitem{DM} N. Deruelle and V.F.
   Mukhanov, ``On matching conditions for cosmological
perturbations",  preprint  gr-qc/9503050.
\bibitem{first}
V. Mukhanov and G. V. Chibisov, JETP Lett. 33,  (1981) 532;\\
A. Guth and S. Y. Pi. Phys. Rev. Lett. 49 (1982) 1110;\\
A.A. Starobinski, Phys. Lett. B117 (1982) 175;\\
S.W. Hawking, Phys. Lett. B115 (1982) 295;\\
J.M. Bardeen, P.S. Steinhardt and M.S. Turner, Phys. Rev. D28 (1983) 679.
\bibitem{MFB} V. Mukhanov, H.A. Feldman and R. Brandenberger,
 Phys. Rep. 215 (1992) 203.
\bibitem{LL} A. R. Liddle and D. H. Lyth, Phys. Rep. 231, (1993) 1.
\bibitem{LIGO} A. Abramovici et al., Science 256 (1992) 325.
\bibitem{VIRGO} B. Caron et. al, ``Status of the VIRGO experiment", preprint
Lapp-Exp-94-15.
\bibitem{thorne} K. S. Thorne, in 300 Years of Gravitation, S. W.
Hawking and W. Israel, Eds. (Cambridge Univ. Press, Cambridge, 1987.
\bibitem{grishchuk} L. P. Grishchuk, Proc. 9th Int. Conf. on General Relativity
and Gravitation, ed. E. Schmutzer, Cambridge Univ. Press, 1983.
 \bibitem{picasso} F. Pegaro, E. Picasso, L. Radicati, J. Phys. A11 (1978)
1949.
\bibitem{caves} C. M. Caves, Phys. Lett B80 (1979) 323.
\bibitem{reece} C. E. Reece et al., Phys. Lett A104 (1984) 341.
\end{thebibliography}
\end{document}